\documentclass[a4paper]{article}

\newcount\commentcount \commentcount=0 
 \long\def\comment#1{\ifnum\commentcount=1 #1\fi}

\newcommand{\printsize}{
      \headheight=0pt                              
     \topmargin=-1cm \headsep=0cm
      \oddsidemargin=-0.5cm \evensidemargin=-0.5cm  
      \textheight=24truecm \textwidth=17truecm      %
	  \setlength{\columnsep}{20pt}                  
}

\newtoks\reportnoregister \newtoks\eprintnoregister
\newcommand{\reportnumber}[1]{\reportnoregister={#1}}
\newcommand{\eprintnumber}[1]{\eprintnoregister={#1}}

\reportnumber{\mbox{}} 
\eprintnumber{\mbox{}} 

\newcommand{\reportid}{
   \begin{minipage}{12cm}\vspace{-2.5cm}  
     \begin{flushright}
      {\normalsize \the\reportnoregister  \\[-.2cm]
       \eprintstyle{\the\eprintnoregister}}\vspace{3.0cm}
     \end{flushright}
   \end{minipage}\hspace{-12cm} }

\catcode`@=11   
\def\title#1{\gdef\@title{\reportid#1}}
\catcode`@=12   

\newcommand{\eprintstyle}[1]{\textsf{#1}} 

\newcommand{\journalfont}{\rm}  
\newcommand{\jou}[1]{{\journalfont #1\ }}
\newcommand{\joudef}[2]{\newcommand #1{\jou{\ignorespaces #2}}}

\joudef{\APAHA}  { Acta Physica Acad.~Sci.~Hung.}
\joudef{\ajp}    { Am.~J.~Phys.}
\joudef{\aaa}    { Astron.\ Astrophys.}
\joudef{\aip}    { Adv.\ Phys.}
\joudef{\adm}    { Adv.\ Math.}
\joudef{\am}     { Ann.\ Math.}
\joudef{\apb}    { Ann.\ Phys.\ (Berlin)}
\joudef{\apny}   { Ann.\ Phys.\ (N.Y.)}
\joudef{\apj}    { Astrophys.\ J.}
\joudef{\apjs}   { Astrophys.\ J.\ Suppl.}
\joudef{\BAPMA}  {Bull.~Acad.~Pol.~Sci.~Ser.~Sci.~Math.~Astron.~Phys.}
\joudef{\baps}   { Bull.~Am.~Phys.~Soc.}
\joudef{\cjp}    { Can.\ J.\ Phys.}
\joudef{\cmda}   { Celest.\ Mech.\ Dyn.\ Astron.}
\joudef{\cmp}    { Commun.\ Math.\ Phys.}
\joudef{\cqg}    { Class.\ Quantum Grav.}
\joudef{\faa}    { Funct.\ Anal.\ Appl.}
\joudef{\grg}    { Gen.\ Rel.\ Grav.}
\joudef{\ijmpd}  { Int.\ J.\ Mod.\ Phys.\ D}
\joudef{\ijtp}   { Int.\ J.\ Theor.\ Phys.}
\joudef{\invm}   { Invent.\ Math.}
\joudef{\jm}     { J.\ Math.}
\joudef{\jmp}    { J.\ Math.\ Phys.}
\joudef{\jpa}    { J.\ Phys.\ A}
\joudef{\jpg}    { J.\ Phys.\ G\relax}
\joudef{\jpamg}  { J.\ Phys.\ A:\ Math.\ Gen.}
\joudef{\jpdap}  { J.\ Phys.\ D:\ Appl.\ Phys.}
\joudef{\lrr}    { Living Rev. Relativity}
\joudef{\mnras}  { Mon.\ Not.\ R.\ Ast.\ Soc.}
\joudef{\mpla}   { Mod.\ Phys.\ Lett.\ A} 
\joudef{\nature} { Nature}
\joudef{\nc}     { Nuovo Cim.}
\joudef{\npb}    { Nuc.\ Phys.\ B}
\joudef{\ph}     { Physica}
\joudef{\pla}    { Phys.\ Lett.\ A}
\joudef{\plb}    { Phys.\ Lett.\ B}
\joudef{\pr}     { Phys.\ Rev.}
\joudef{\pra}    { Phys.\ Rev.\ A}
\joudef{\prb}    { Phys.\ Rev.\ B}
\joudef{\prc}    { Phys.\ Rev.\ C}
\joudef{\prd}    { Phys.\ Rev.\ D}
\joudef{\prep}   { Phys.\ Rep.}
\joudef{\prl}    { Phys.\ Rev.\ Lett.}
\joudef{\PANUE}  {Phys.~Atom.~Nucl.}
\joudef{\pnas}   { Proc.\ Natl.\ Acad.\ Sci.\ USA}
\joudef{\prsla}  { Proc.\ Roy.\ Soc.\ Lond.\ A}
\joudef{\ptp}    { Prog.\ Theor.\ Phys.}
\joudef{\ptps}   { Prog.\ Theor.\ Phys.\ Suppl.}
\joudef\rmp      { Rev.\ Mod.\ Phys.}
\joudef\spj      { Sov.\ Phys.\ JETP}
\joudef\jetpl    { JETP Lett.}

\newcommand{\approxmetheqmotion}{04.25.-g}  
\newcommand{\physicsblackholes}{04.70.-s}  

\catcode`@=11

\newcommand\eqalign[1]{\null\,\vcenter{\openup\jot\m@th
  \ialign{\strut\hfil$\displaystyle{##}$&$\displaystyle{{}##}$\hfil
      \crcr#1\crcr}}\,}
\newcommand\meqalign[1]{\null\,\vcenter{\openup\jot\m@th
  \ialign{\strut\hfil$\displaystyle{##}$&&$\displaystyle{{}##}$\hfil
      \crcr#1\crcr}}\,}
\def\ps@reportnumber{%
    \let\@oddfoot\@empty\let\@evenfoot\@empty
    \def\@oddhead{\hfil\rightmark}}
	
\catcode`@=12   

\newdimen\arrayruleHwidth
\makeatletter
\newcommand\Hline{\noalign{\ifnum0=`}\fi\hrule \@height \arrayruleHwidth
  \futurelet \@tempa\@xhline}
\makeatother

\newcommand\thickbaselines{\baselineskip=20pt\lineskip=3pt\lineskiplimit=3pt}

\catcode`@=11

\renewcommand\matrix[1]{\null\,\vcenter{\thickbaselines\m@th
    \ialign{\hfil$##$\hfil&&\quad\hfil$##$\hfil\crcr
      \mathstrut\crcr\noalign{\kern-\baselineskip}
      #1\crcr\mathstrut\crcr\noalign{\kern-\baselineskip}}}\,} 
\catcode`@=12   

\renewcommand{\d}{{\rm d}} 

 \newcommand\Hscr{{\cal H}}

\newcommand\undersim[1]{\mathop{\vtop{\ialign{##\crcr
     $\hfil\displaystyle{#1}\hfil$\crcr\noalign
     {\kern1pt\nointerlineskip}\hbox{$\hfil\sim\hfil$}\crcr
     \noalign{\kern1pt}}}}}

\printsize

\usepackage{amsmath}

\newcommand\dottheta{\,\dot{\!\theta}}

\begin{document}
\bibliographystyle{prsty}



\title{
CARTER'S CONSTANT REVEALED}
\vspace{20pt}
\author{K. Rosquist$^{*\dag}$, T. Bylund$^*$, L. Samuelsson$^\ddag$\\[10pt]
\emph{\normalsize *Department of Physics, AlbaNova University Center} \\[-5pt]
\emph{\normalsize Stockholm University},
\emph{\normalsize 106 91 Stockholm, Sweden} \\[5pt]
\emph{\normalsize ${}^\dag$ICRANet, Piazzale della Repubblica, 10}\\[-5pt]
\emph{\normalsize 65100 Pescara, Italy} \\[5pt]
\emph{\normalsize {}$^\ddag$Nordita, AlbaNova University Center} \\[-5pt]
\emph{\normalsize 106 91 Stockholm, Sweden} \\[1.3cm]
%
\begin{minipage}[t]{0.8\linewidth}\small{
A new formulation of Carter's constant for geodesic motion in Kerr black holes is given. It is shown that Carter's constant corresponds to the total angular momentum plus a precisely defined part which is quadratic in the linear momenta. The characterization is exact in the weak field limit obtained by letting the gravitational constant go to zero. It is suggested that the new form can be useful in current studies of the dynamics of extreme mass ratio inspiral (EMRI) systems emitting gravitational radiation.
}
\end{minipage}}

\date{}

\maketitle

\centerline{\bigskip\noindent
PACS numbers: \physicsblackholes, \approxmetheqmotion}


\section{INTRODUCTION}
The Kerr solution together with its charged counterpart represents one of the most important advances in general relativity since its original formulation by Einstein almost a century ago. As a stationary axisymmetric geometry it admits one timelike and one spacelike Killing vector field. The geodesic motion therefore has two constants of the motion which are linear in the momenta corresponding to conservation of energy and conservation of the azimuthal angular momentum.  As always, there is also the quadratic constant of the motion corresponding to conservation of mass. 

Carter found an additional nontrivial quadratic constant of the motion \cite{Carter:1968}. Ever since its discovery almost four decades ago, there has been an ongoing discussion about the physical interpretation of Carter's constant, see e.g.\ \cite{de_Felice:1980, de_Felice&Preti:1999, de_Felice&Preti:2000, Faridi:1986}. The constant is important when calculating the spectrum of gravitational waves emanating from the dynamics of extreme mass ratio inspiral (EMRI) type systems in particular (see \cite{Glampedakis:2005} for a review). The consensus in the existing literature has been that Carter's constant represents a generalization of the total squared angular momentum. Indeed, in the limit of zero rotation $a \rightarrow 0$ it does reduce to the total angular momentum (minus the conserved azimuthal part). However, there has also a been a consensus that it is not exactly the total angular momentum, but that it corresponds to the total angular momentum plus an additional part. In this note, we give a precise and unambiguous characterization of Carter's constant. In addition to the zero angular momentum limit, $a \rightarrow 0$, there is also another limit which is physically relevant. This is the weak field limit $G \rightarrow 0$ in which Newton's gravitational constant tends to zero. For example, the azimuthal angular momentum $L_z$ is the same in both these limits. However, it turns out that Carter's constant does not reduce to a pure angular momentum in the weak field limit. Instead, there are additional terms quadratic in the linear momenta which are proportional to the square of $a$, the angular momentum parameter in the Kerr geometry. Below, we give a more detailed description and interpretation of this result.

The limit $G \rightarrow 0$ is mathematically equivalent to the limit $M \rightarrow 0$ corresponding to the source mass tending to zero since $G$ only occurs in the the combination $GM$ in the Kerr solution. From the physical point of view however, the limit $M \rightarrow 0$ (while keeping $a$ fixed) can only be taken, strictly speaking, for overextreme systems ($M<a$), which hence do not include black holes. However, it is reasonable to expect that the physical interpretation of Carter's constant should in principle be the same for both underextreme and overextreme systems. Mathematically, the limit $M \rightarrow 0$ is well defined for any Kerr solution. Besides, overextreme astrophysical systems are interesting in themselves as they are presumably not uncommon. Notably, they include the solar system for which $a/M \approx 40$ \cite{Rosquist:2008a} where $a=J/M$ and $J$ is the total angular momentum. It also worth noting that for charged black holes, the limits $G \rightarrow 0$ and $M \rightarrow 0$ are not equivalent.

\section{THE IDEA}
In order to formulate our basic idea, it is useful to express the Kerr metric in terms of a suitable Minkowski frame $M^\mu$ meaning that $\eta= \eta_{\mu\nu} M^\mu M^\nu$ is the Minkowski metric. We use relativistic units with $c=1$, but keep $G$ explicit. Then we define a Minkowski frame expressed in Boyer-Lindquist coordinates by
\begin{equation}
   M^0 = \frac{\rho_0}{\rho}(\d t-a\sin^2\!\theta \d\phi) \ ,\quad
   M^1 = \frac{\rho}{\rho_0} \d r \ ,\quad
   M^2 = \rho\d\theta \ ,\quad
   M^3 = \frac{\sin\theta}{\rho}(-a\d t+\rho_0^2\d\phi)
\end{equation}
where the functions $\rho$ and $\rho_0$ are given by
\begin{equation}
     \rho^2 = r^2 + a^2 \cos^2\!\theta \ ,\qquad \rho_0^2 = r^2 + a^2 \ .
\end{equation}
This gives the Minkowski metric in oblate spheroidal coordinates. They are related to Cartesian coordinates by
\begin{equation}
 \begin{split}
  x &= \rho_0 \sin\theta \cos\phi \\
  y &= \rho_0 \sin\theta \sin\phi \\
  z &= r \cos\theta \ .
 \end{split}   
\end{equation}
The explicit form of the Minkowski metric in these coordinates is
\begin{equation}\label{Mspheroidal}
   \eta = \tilde\eta_{\mu\nu} \d \tilde x^\mu \d \tilde x^\nu
        = -\d t^2 + \frac{\rho^2}{\rho_0^2} \d r^2 + \rho^2 \d\theta^2
         + \rho_0^2 \sin^2\!\theta \d\phi^2 \ ,
\end{equation}
where $\tilde x^\mu = (t,r,\theta,\phi)$ denotes the spheroidal coordinates. The Kerr metric itself is
\begin{equation}\label{Kerr}
   g = -h(M^0)^2 + h^{-1}(M^1)^2 + (M^2)^2 + (M^3)^2 \ ,
\end{equation}
where
\begin{equation}
   h = 1-\frac{2GMr}{r^2+a^2} \ .
\end{equation}
Note that the Minkowski frame $M^\mu$ does not contain the
mass\footnote{The parameter $a$ has the interpretation of angular
momentum per unit mass. Since the angular momentum itself scales
linearly with the mass, $a$ should be considered as a parameter which
does not depend on the mass.} which appears only in the function
$h$. In addition, the function $h$ depends on only one of the
coordinates, $r$. Taking the weak field limit $G \rightarrow 0$ of the
Kerr metric \eqref{Kerr} obviously gives back Minkowski spacetime in
oblate spheroidal coordinates.
Carter's constant is given by \cite{Misner_etal:1973}
\begin{equation}
   Q = p_\theta^2 + p_\phi^2 \cot^2\!\theta - a^2(p_t^2+2\Hscr)\cos^2\!\theta
\end{equation}
where the Hamiltonian $\Hscr$ is related to the mass $\mu$ of the particle by
\begin{equation}
   \Hscr = \tfrac12 g^{\mu\nu}p_\mu p_\nu = -\tfrac12 \mu^2 \ .
\end{equation}
Using this relation and writing $E=-p_t$, $L_z= p_\phi$ for the energy
and azimuthal angular momentum, the constant can be rewritten as
\begin{equation}\label{Carter2}
   Q = p_\theta^2 + L_z^2 \cot^2\!\theta - a^2(E^2-\mu^2)\cos^2\!\theta \ .
\end{equation}
In the limit of zero angular momentum of the source, $a \rightarrow
0$, we have as is well-known
\begin{equation}\label{Qlimit}
   Q = p_\theta^2 + L_z^2 \cot^2\!\theta = L_x^2 + L_y^2 = L^2 - L_z^2 
\end{equation}
where
\begin{equation}\label{L2spherical}
   L^2 = p_\theta^2 + \frac{p_\phi^2}{\sin^2\!\theta}
\end{equation}
is the total squared angular momentum expressed in spherical
coordinates and $L_x$ and $L_y$ are components of the angular
momentum corresponding to equatorial symmetry axes.

In order to extend this result, we make the crucial observation that
Carter's constant as given in \eqref{Carter2} does not contain the
source mass $M$ explicitly. It contains only the parameters $L_z$, $E$
and $\mu$ which refer to the particle. Therefore the expression in
\eqref{Carter2} must itself be a constant of the motion in flat
space. At this point it is essential that oblate spheroidal
coordinates are used in \eqref{Carter2}. We can now reinterpret the
expression for $Q$ in \eqref{Carter2} using flat space spheroidal
coordinates by setting $E=-p_t$ and $\mu^2 = -2 \Hscr_{\text{flat}} =
-\tilde\eta^{\mu\nu} p_\mu p_\nu$ where $\tilde\eta^{\mu\nu}$ are the
contravariant components of the metric \eqref{Mspheroidal}. This gives
\begin{equation}\label{Qflat}
   Q_0 = \frac{r^2}{\rho^2}p_\theta^2
    +\frac{ r^2\cot\!^2\theta}{\rho_0^2}p_\phi^2 
    - \frac{a^2 \rho_0^2 \cos^2\!\theta}{\rho^2} p_r^2 \ .
\end{equation}
Where the subscript ``0'' is a reminder that this expression is valid in
flat space only.  It must necessarily correspond to some
combination of the flat space constants of the motion. Guided by \eqref{Qlimit}, we first try with $L_x^2 + L_y^2$. To express that constant in spheroidal coordinates we note that the total angular momentum is
\begin{equation}\label{L2spheroidal}
   L^2 = \frac{\rho_0^2}{\rho^4}\left(rp_\theta - a^2\sin\theta\cos\theta p_r\right)^2 +
         \left(1-\frac{a^2\cos^2\theta}{\rho_0^2}\right)\frac{p_\phi^2}{\sin^2\!\theta}\ .
\end{equation}
In the limit $a \rightarrow 0$, this reduces to \eqref{L2spherical} as
it should. Subtracting $L_z^2 = p_\phi^2$ changes only the second term
and we obtain
\begin{equation}\label{Leq}
   L_x^2 + L_y^2 = L^2 - L_z^2
       = \frac{\rho_0^2}{\rho^4}\left(rp_\theta - a^2\sin\theta\cos\theta p_r\right)^2 + 
         \frac{r^2\cot^2\!\theta}{\rho_0^2}p_\phi^2 \ .
\end{equation}
The flat space version of Carter's constant in \eqref{Qflat} has no cross terms in the momenta whereas the above expression for $L_x^2 + L_y^2$ contains a cross term in $p_rp_\theta$. Therefore we need to add some other (flat space) constant of the motion which can cancel this cross term. It turns out that $p_z^2$ is exactly what is needed to do that job. Expressing $p_z$ in oblate spheroidal coordinates gives
\begin{equation}\label{pz}
   p_z = \frac{\rho_0^2\cos\theta}{\rho^2}p_r
          -\frac{r\sin\theta}{\rho^2}p_\theta \ .
\end{equation}
Finally combining the expressions \eqref{Qflat}, \eqref{Leq} and
\eqref{pz} gives the amazingly simple relation
\begin{equation}\label{Qconst}
   Q_0 = L_x^2 + L_y^2 - a^2 p_z^2 \ .
\end{equation}
This is our main result. For future reference we note that the (flat space) total angular momentum can be written in the form
\begin{equation}\label{L2flat}
   L^2 = Q_0 + L_z^2 + a^2 p_z^2 \ .
\end{equation}
We also define the related positive definite constant of the motion
\begin{equation}
   P = Q + a^2 E^2 = p_\theta^2 + L_z^2 \cot^2\!\theta 
                    + a^2 E^2 \sin^2\!\theta + a^2 \mu^2 \cos^2\!\theta \ .
\end{equation}
Its flat space form is
\begin{equation} \label{QEconst}
   P_0 = L_x^2 + L_y^2 + a^2(p_x^2 + p_y^2) + a^2 \mu^2 \ .
\end{equation}

\section{INTERPRETATION}
In this section we make some comments on the new characterization of
Carter's constant given above. We begin by focusing attention on the
form \eqref{Qconst}. It is then clear that the equatorial projection
of the angular momentum and the momentum in the $z$-direction have a
tendency to follow each other. This can be regarded as natural from
the physical point of view, since both quantities are related to
motion off the equatorial plane. Looking instead at the form
\eqref{QEconst} shows that the equatorially projected angular momentum
has a tendency to oppose horizontal motion and vice versa. There is
therefore a certain jet-like effect in that horizontal motion is
disfavored unless it occurs in the equatorial plane itself.
Based on the results discussed above, in particular equation
\eqref{L2flat}, one may define a total (non-conserved) angular
momentum for a particle in the Kerr geometry by
\begin{equation}\label{L2def}
   \tilde{L}^2 := Q + L_z^2 + a^2 p_z^2 \ ,
\end{equation}
where $p_z$ is given by \eqref{pz} and the momenta are defined with
the Kerr metric factor $h$ included, $p_r = h^{-1} \rho_0^{-2} \rho^2
\dot r$, $p_\theta= \rho^2 \dottheta$. Referring to \eqref{pz}, the
explicit form of the Kerr $p_z$ momentum is therefore given by
\begin{equation}
   p_z = h^{-1} \cos\theta \, \dot r - r \sin\theta \, \dottheta \ .  
\end{equation}
It is clear that $\tilde L^2$ reduces to the conserved total angular momentum
in both limits, $a \rightarrow 0$ and $G \rightarrow 0$. From its
definition it is also evident that $\tilde L^2$ differs from the flat space
angular momentum only by factors of $h$. We therefore expect that
$\tilde L^2$ is approximately constant in the region $\tilde h \ll1$ where
$\tilde h=1-h = 2GMr/(r^2+a^2)$. Since the horizon is defined by
$h=0$, this corresponds to the region well outside the horizon, $r \gg
r_{h}$.

Glampedakis et al.\ \cite{Glampedakis_etal:2002} define a ``spherical"
total angular momentum by
\begin{equation}\label{L2GHK}
   \tilde L^2_{\text{GHK}} = p_\theta^2 + \frac{p_\phi^2}{\sin^2\!\theta} \ .
\end{equation}
Although formally identical to the expression \eqref{L2spherical}, it
should be borne in mind that \eqref{L2GHK} is an expression in
spheroidal coordinates, not spherical. Therefore, although the
Glampedakis et al.\ angular momentum function $\tilde L_\text{GHK}^2$ does reduce
to the conserved angular momentum in the limit $a \rightarrow 0$, it
does not in the limit $G \rightarrow 0$, for which the correct
expression should be \eqref{L2spheroidal}. 

Our point of view is that since angular momenta should be related to  
(spatial) rotational symmetries, the only angular momentum which can  
be defined for the Kerr metric is $L_z = p_\phi$. Only by taking  
appropriate limits (such as $a \rightarrow 0$ and $G \rightarrow 0$)  
can one achieve a setting in which also other angular momenta can be  
defined. Comparing with the work of de Felice and Preti  
\cite{de_Felice&Preti:1999}, they regard the spherical-like expression  
\eqref{L2GHK} as an angular momentum in their attempts to give a  
physical meaning to Carter's constant. However, referring to the above  
discussion, such an interpretation does not have an invariant meaning.  
This is most certainly the reason why their additional term depends on  
$p_r$ rather than $p_z$ as obtained in our analysis.

Finally we may speculate that the insights gained in this work may
prove practically useful in e.g.\ improving approximate methods for
calculating gravitational waveforms from EMRIs, see e.g.\
\cite{Glampedakis:2005}. In particular it would be interesting to
check whether our $\tilde{L}^2$ is even more close to being conserved
than $\tilde{L}_{\text{GHK}}^2$.

%

\bibliography{kr}

\begin{thebibliography}{1}

\bibitem{Carter:1968}
B. Carter, \pr {\bf 174},  1559  (1968).

\bibitem{de_Felice:1980}
F. de~Felice, \jpamg {\bf 13},  1701  (1980).

\bibitem{de_Felice&Preti:1999}
F. de~Felice and G. Preti, \cqg {\bf 16},  2929  (1999).

\bibitem{de_Felice&Preti:2000}
F. de~Felice and G. Preti, \jpamg {\bf 33},  2767  (2000).

\bibitem{Faridi:1986}
A.~M. Faridi, \grg {\bf 18},  271  (1986).

\bibitem{Glampedakis:2005}
K. Glampedakis, \cqg {\bf 22},  S605  (2005), \onlineversion{gr-qc/0509024}.

\bibitem{Rosquist:2008a}
K. Rosquist,  in {\em Proc.\ Eleventh Marcel Grossmann Meeting on General
  Relativity}, edited by H. Kleinert, R. Jantzen, and R. Ruffini (World
  Scientific, Singapore, 2008), p.\ 2294.

\bibitem{Misner_etal:1973}
C.~W. Misner, K.~S. Thorne, and J.~A. Wheeler, {\em Gravitation} (Freeman, San
  Francisco, USA, 1973).

\bibitem{Glampedakis_etal:2002}
K. Glampedakis, S.~A. Hughes, and D. Kennefick, \prd {\bf 66},  064005  (2002).

\end{thebibliography}
\end{document}